\documentclass[11pt,preprint]{aastex}
\usepackage{lscape}

\def\kms{km~s$^{-1}$}

\begin{document}

\title{
The chemical composition of NGC~5824, a globular cluster without iron spread but 
with an extreme Mg-Al anticorrelation
\footnote{Based on observations collected at the ESO-VLT under the program
095.D-0290.}}

\author{Alessio Mucciarelli$^{2,3}$,
Emilio Lapenna$^{2.3}$,
Francesco R. Ferraro$^{2,3}$,
Barbara Lanzoni$^{2,3}$}

\affil{$^{2}$Dipartimento di Fisica \& Astronomia, Universit\`a 
degli Studi di Bologna, via Gobetti 93/2, I-40129,
Bologna, Italy}

\affil{$^{3}$INAF - Osservatorio di Astrofisica e Scienza dello Spazio, 
via Gobetti 93/3, I-40129, Bologna, Italy}

\begin{abstract}

NGC~5824 is a massive Galactic globular cluster suspected to have an intrinsic spread in 
its iron content, according to the strength of the calcium triplet lines. 
We present chemical abundances of 117 cluster giant stars using high-resolution spectra 
acquired with the multi-object spectrograph FLAMES. 
The metallicity distribution of 87 red giant branch stars is peaked at [Fe/H]=--2.11$\pm$0.01
dex, while that derived from 30 asymptotic giant branch stars is peaked at
[Fe/H]=--2.20$\pm$0.01 dex. Both the distributions are compatible with a null spread, 
pointing out that this cluster did not retain the ejecta of supernovae. 
The small iron abundance offset between the two groups of stars is similar to those already observed 
among red and asymptotic giant branch stars in other clusters.
The lack of intrinsic iron spread rules out the possibility that NGC~5824 is
the remnant of a disrupted dwarf galaxy, as previously suggested. 
We also find evidence of the chemical anomalies usually observed in globular clusters, 
namely the Na-O and the Mg-Al anticorrelations. In particular, NGC~5824 exhibits 
a huge range of [Mg/Fe] abundance, observed only in a few metal-poor and/or massive clusters.
We conclude that NGC~5824 is a normal globular cluster, without spread in [Fe/H] 
but with an unusually large spread in [Mg/Fe], possibly due to an efficient self-enrichment 
driven by massive asymptotic giant branch stars.
\end{abstract}



\section{Introduction}

The majority of the Galactic globular clusters (GCs) studied so far 
through high-resolution spectroscopy reveals two chemical signatures, 
usually considered as the key features to define a stellar system as a GC: 
(1)~a very small star-to-star scatter in their iron abundance, compatible, 
within their uncertainties, with a null spread \citep[see e.g.][]{carretta09,willman12};
(2)~the presence of star-to-star variations in the chemical abundances of some
light elements, structured in some well-defined patterns, like the Na-O and Mg-Al 
anticorrelations \citep[see e.g.][]{carretta09g,carretta09u,meszaros15,pancino17}.\\
The first evidence points out that GCs were not massive enough to retain in their 
gravitational well the high-velocity ejecta of the supernovae. The second evidence instead 
is interpreted as the signature of the retaining/recycling of the low-velocity ejecta 
of some polluter stars where high-temperature proton-capture cycles (CNO, NeNa and MgAl chains) occurred.
While the details of this self-enrichment process are still unclear and under debate 
\citep[see e.g. the critical discussions by \citet{bastian15} and][]{renzini15}, 
the presence of chemically distinct stellar populations in GCs is now widely recognized and accepted.

While the Na-O anticorrelation has been detected in almost all old GCs studied so far 
(at least in those with statistically significant samples
\footnote{ We remind the case of Ruprecht 106 that does not 
show chemical anomalies in the light elements from the analysis of 9 stars \citep{villanova13}.}), 
some undeniable exceptions to the first evidence quoted 
above are known. There are three massive stellar systems, usually labelled 
as GCs according to their appearance and brightness profile, that show large metallicity distributions: 
Omega Centauri in the Galactic Halo \citep{johnson10,pancino11,marino11a}, 
Terzan~5 in the Galactic Bulge \citep{ferraro09,origlia11,origlia13,massari14}
and M54 in the Sagittarius dwarf galaxy \citep{brown99,bellazzini08,carretta10a,carretta10b,mu17}. 
The first two stellar systems have multi-modal [Fe/H] distributions, covering 
a range of $\sim$1 dex, while the metallicity distribution of M54,
once the contamination of the Sagittarius stars has been taken into account, is significantly 
smaller than those of Omega Centauri and Terzan~5.
Even if characterized by the largest metallicity distributions observed so far among 
the GC-like systems, these three systems cannot be easily explained within the same 
framework: 
Omega Centauri is usually interpreted as the remnant of a disrupted dwarf spheroidal 
galaxy \citep{bekki03}, 
Terzan~5 is likely the fossil relic of one of the primordial structures that 
contributed to build up the Galactic Bulge \citep{ferraro16}, while M54, due to 
its position coincident with the center of the Sagittarius galaxy and the strong difference between 
its dispersion velocity profile and that of Sagittarius, should be a GC formed 
independently by the true nucleus of the galaxy and decayed to the present-day position due to 
dynamical friction \citep[as demonstrated by][]{bellazzini08}.

Recently, other GCs have been proposed  to have a intrinsic iron spread (but smaller than 
those observed in three stellar systems quoted above). 
Small intrinsic [Fe/H] scatters (of the order of 0.1 dex), based on both low- and high-resolution spectroscopy, have been claimed 
for M~22 \citep{marino09}, NGC~3201 \citep{simmerer13}, M~2 \citep{yong14}, 
NGC~5824 \citep[][hereafter DC14]{dacosta14}, NGC~5286 \citep{marino15} and 
M~19 \citep{johnson16}.
For most of these clusters (like M~2, M~22, NGC~5286 and M~19), 
spreads in the s-process and C+N+O abundances (with [s/Fe] and C+N+O abundance 
ratios increasing with [Fe/H]) has been detected, similar to that observed in Omega Centauri 
(no information about s-process element abundances
are available so far for Terzan~5 and M54\footnote{Note that \citet{carretta14} quoted the average 
[s/Fe] abundance ratios measured in the UVES stars of M54 already discussed in 
\citet{carretta10a,carretta10b} but not the abundances for the individual stars. 
This makes impossible to understand whether also M54 shows the same behavior of [s/Fe] 
with [Fe/H].}).
The similarities (but at lower extent) with Omega Centauri have suggested that these clusters 
could be the remnant of dwarf galaxies accreted from our Galaxy
\citep[see e.g. the discussion in][]{marino15}.

Even if fascinating, this hypothesis needs to be supported from extensive and robust spectroscopic 
studies, because several different effects can mimic a spurious iron spread. 
In the case of NGC~3201, the iron spread originally proposed by \citet{simmerer13} was 
found to be due to the inclusion in the sample of some AGB stars that show systematic under-abundances 
of [FeI/H] \citep[see][]{lp14} and hence erroneously interpreted as a metal-poor tail in the cluster 
metallicity distribution \citep{mu15}. 
In the remote GC NGC~2419 (the GC with the largest [Mg/Fe] spread, having stars 
with [Mg/Fe] extended down to $\sim$--1 dex) the strength of the Ca~II triplet lines 
provides a wide metallicity distribution \citep[see][]{ibata11}. 
However, this iron spread is artificial because the strength of the Ca~II triplet lines 
(at a constant Fe and Ca abundance) significantly increases in those stars characterized 
by a relevant depletion in Mg, being Mg one of the most important electron donors in the giant star 
atmospheres \citep{mu12}.
Also the case of M22 has been revised by \citet{mu15b} demonstrating that the use of 
photometric gravities (instead of the spectroscopic ones) leads to 
a narrow iron distribution, compatible with a null intrinsic spread, when FeII lines are measured 
(while the use of spectroscopic gravities provides two broad metallicity distributions 
but also implies unrealistic and too low stellar masses). A similar approach has been applied by \citet{lardo16} to M2, 
thus reducing the iron spread claimed for this cluster by \citet{yong14}.

In this paper we focus our attention on the outer halo GC NGC~5824, another cluster 
for which a modest iron spread was claimed.
First, \citet{saviane12} suggested a possible metallicity spread based on metallicity 
inferred from the strength of the Ca~II triplet lines of 17 stars, subsequently confirmed by DC14
adopting the same technique for a sample of 108 RGB cluster stars. 
The [Fe/H] distribution derived by DC14 has an 
average value of --2.01$\pm$0.01 dex with an observed spread of 0.07 dex.
DC14 proposed that also NGC~5824 could be another case of remnant of a disrupted dwarf galaxy.

Using the Magellan II Telescope, \citet[][hereafter Ro16]{roederer16}  measured the chemical 
composition of 26 stars using the Michigan/Magellan Fiber System (with a spectral resolution 
of $\sim$34000 and a spectral coverage of 4425-4635 \AA\ ) and of two stars using the 
MIKE spectrograph (spectral resolution of $\sim$40000 and coverage of $\sim$3350-9150 \AA\ ). 
The total sample provides an average [Fe/H] abundance of 
--2.38$\pm$0.01 dex ($\sigma$=~0.08 dex) from FeI lines and 
of --1.94$\pm$0.02 dex ($\sigma$=~0.08 dex) from FeII lines. 
The observed dispersions from both the two iron distributions are compatible 
with a null spread, even if Ro16 noted that their observed targets are 
in the brightest portion of the RGB where the corresponding targets by DC14 
provide a small dispersion (while a larger dispersion is found for the faintest stars). 
Ro16 concluded that their high-resolution 
spectroscopic sample could be not adequate to provide a conclusive answer 
about a possible iron spread in this cluster.

In this paper we investigate the chemical composition of NGC~5824 using a sample 
of high-resolution spectra collected with FLAMES at the 
Very Large Telescope for a total of 117 member stars..

\section{Observations}
The observations have been collected under the program 095.D-0290 (PI: Mucciarelli) and 
performed during the night May 28 2015 
with the multi-object spectrograph FLAMES \citep{pasquini} at the Very Large Telescope 
of ESO in the GIRAFFE+UVES combined mode, allowing the simultaneous allocation of  
132 mid-resolution GIRAFFE fibers and 8 high-resolution UVES \citep{dekker00} fibers.
The adopted setups are GIRAFFE HR21 (with a spectral coverage of $\sim$8480--9000 \AA\  
and a spectral resolution of 18000) 
and UVES Red Arm 580 (with a spectral coverage of $\sim$4800--6800 \AA\  and a spectral resolution of 45000).
Two configurations of targets have been defined, 
the first one has been observed with 4 exposures 
of 45 min each and the second one with 5 exposures of 45 min each.
The targets have been selected from the WFPC2@HST photometric catalog 
by \citet{sanna14}, picking only giant stars predicted to be not contaminated 
within the FLAMES fiber diameter by neighbor stars with brighter or comparable magnitudes. 
In the target allocation procedure the highest priority has been attributed to stars 
already observed: the UVES fibers were allocated to stars already observed by DC14 and Ro16, and the 
GIRAFFE fibers to the remaining DC14 targets for which the membership has been already established.
Most of the targets have been allocated within $\sim$350 arcsec from the cluster 
center; at larger radii the color-magnitude diagram (CMD) is dominated by field stars 
and the cluster sequences are barely recognizable.
The residual fibers have been allocated to giant stars in the external regions, 
in order to have a sample of surrounding field stars for the identification of possible 
Galactic interlopers among the stars observed close to the cluster.
A total of 211 stars have been observed (205 with GIRAFFE and 6 with UVES). 
88 of them are in common with DC14, while 23 are in common 
with Ro16 (all in common with DC14).
Finally, about 20 GIRAFFE fibers and two UVES fibers 
have been dedicated to observe empty sky regions in order to sample the 
sky background.

The spectra have been reduced using the dedicated ESO pipelines that 
perform bias-subtraction, flat-fielding, wavelength calibration, 
one-dimensional spectral extraction and (only for the UVES spectra) 
order merging. 
For each exposure, the spectra of the sky regions have been 
combined together and the derived master-sky spectrum subtracted to each individual 
stellar spectrum.  The latter have been corrected for the corresponding 
heliocentric radial velocity (RV) as explained in Section \ref{rv} and finally
the spectra of each target coadded together, reaching signal-to-noise ratio (SNR) per pixel 
from $\sim$70 for the faintest GIRAFFE targets (V$\sim$17.8) 
to $\sim$260 for the brightest ones (V$\sim$15.4). 
For the UVES spectra a SNR per pixel of $\sim$50 is reached.


\section{Radial velocities}
\label{rv}
RVs have been measured adopting the standard cross-correlation 
method as implemented in the {\tt IRAF} task {\tt FXCOR}. 
As template spectrum we adopted a synthetic spectrum calculated 
with the {\tt SYNTHE} code \citep[see][]{sbordone04} and convoluted with a Gaussian profile 
in order to reproduce the instrumental profile.
Before to measure the RV on the photospheric lines, we checked 
the correctness of the wavelength calibration by measuring the 
position of some sky emission lines: both for GIRAFFE and UVES spectra 
no significant offset in the zero-point of the wavelength calibration is found. 

For each target, the measure of RV has been performed on the spectra of  individual exposures
and then the final heliocentric RV has ben computed as average of the 
individual values. This approach allows to detect possible binary stars, 
at least those stars with RV variations detectable within the same night.
We identified only one star (namely \#29580) that shows a dispersion of the average RV significantly 
larger than those measured in stars of similar magnitude. 
This star, even if likely cluster member according to its median RV and the 
small distance from the cluster center ($\sim$60 arcsec), has been excluded
from the following chemical analysis.

Table~\ref{info1} lists for all the targets the final heliocentric RVs computed 
as non-weighted mean of the individual measures; the quoted uncertainties 
have been computed as the dispersion of the mean normalized to the 
root mean square of the number of used exposures.

For the 88 targets in common with DC14 
an average RV difference (in the sense our study - DC14) 
of +2.5$\pm$1.1 \kms ($\sigma=$10.1 \kms) is found, 
where the dispersion is dominated by the uncertainties in the DC14 RVs, 
due to the lower spectral resolution (R$\sim$2500).
For the 23 stars in common with Ro16 we found 
+0.7$\pm$0.2 \kms ($\sigma=$0.8 \kms).

\begin{figure*}
\plotone{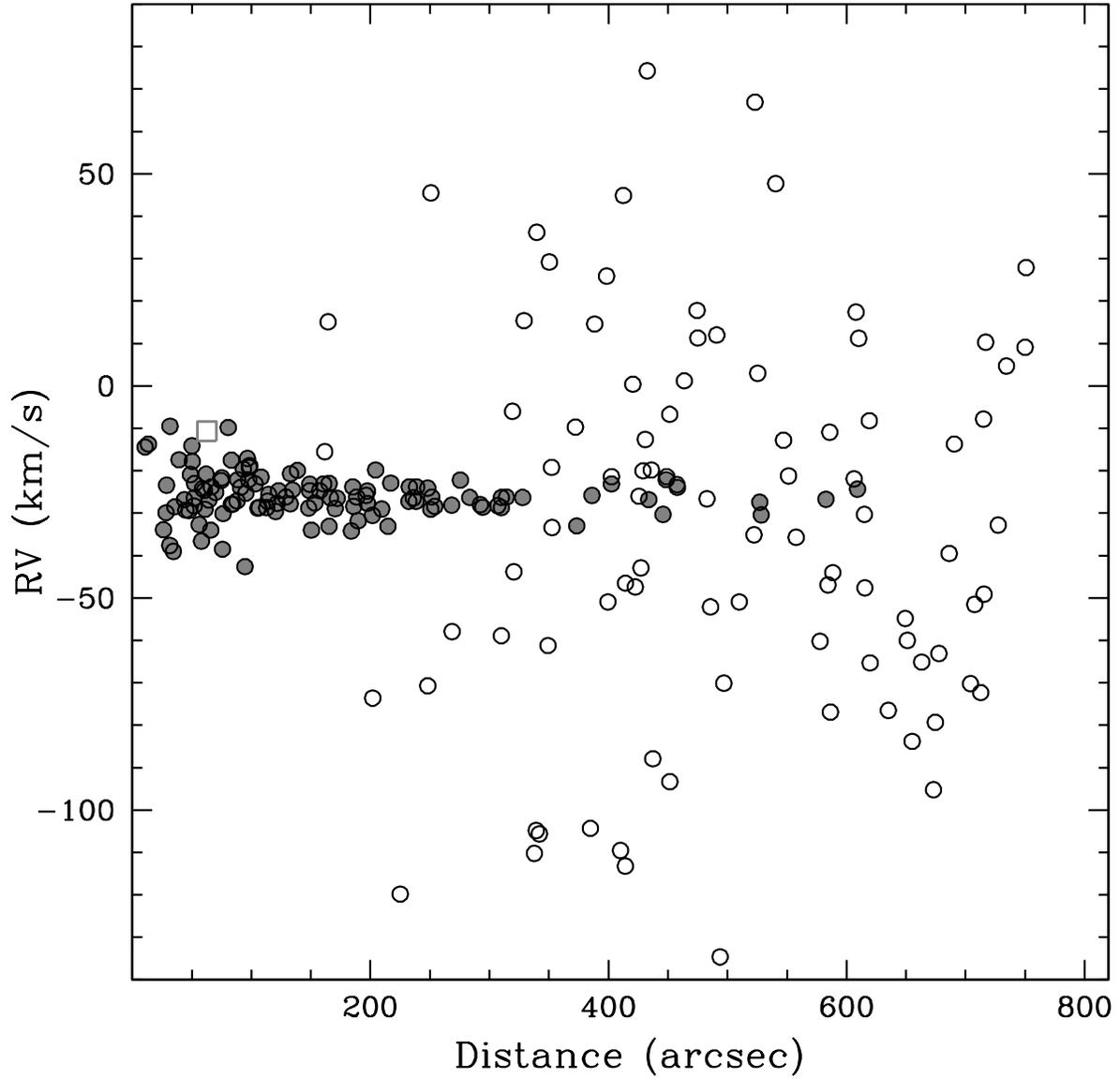}
\caption{Behaviour of RV as a function of the distance from the cluster center for all the observed targets 
of NGC~5824. Grey filled points are the stars flagged as cluster members according to their 
RV and [Fe/H]. Grey empty square indicates a candidate binary star.}
\label{rvd}
\end{figure*}

Fig.~\ref{rvd} shows the behavior of the heliocentric RVs of all the targets
as a function of the distance from the cluster center 
quoted by \citet{sanna14}. As visible, a group of stars 
with RV clumped around $\sim$--26 \kms is clearly recognizable
until $\sim$350 arcsec, while for larger radii the sample 
shows a significant RV dispersion. 
The grey points are the stars selected as member cluster stars 
according to their RV and [Fe/H] (see Section~\ref{members}).
The position of the candidate binary star is shown as a square symbol.

\section{Atmospheric parameters}
\label{atmos}
The determination of the atmospheric parameters has been performed using photometric information.
The magnitudes of the catalog by \citet{sanna14} have been reported in standard Johnson photometric 
system using the stars in common with the catalog of photometric standard stars 
by P. B. Stetson\footnote{http://www.cadc-ccda.hia-iha.nrc-cnrc.gc.ca/en/community/STETSON/standards/}. 
Effective temperatures ($T_{\rm eff}$) have been computed by means the $(U-V)_0$--$T_{\rm eff}$ 
transformation by \citet{alonso99,alonso01} and adopting the color excess E(B-V)=~0.14 mag 
\citep{ferraro99}.
The extinction coefficients are from \citet{mccall04}.
Because of the dependence on the metallicity of the $(U-V)_0$--$T_{\rm eff}$ relation, 
we adopted as guess value a metallicity [Fe/H]=--2.0 dex (according to DC14 and Ro16) 
for all the targets. After a first determination of the chemical abundances, $T_{\rm eff}$ 
have been refined using the proper metallicity of each star. 

Surface gravities (log~g) have been estimated using the Stefan-Boltzmann relation, assuming 
the photometric $T_{\rm eff}$, the bolometric corrections estimated according to \citet{alonso99} 
and the true distance modulus quoted by \citet{ferraro99} ($(m-M)_0$=~17.53 mag).
In the first determination of log~g, we assumed a stellar mass of 0.75 $M_{\odot}$; 
subsequently we attributed a stellar mass of 0.65 $M_{\odot}$ for the member stars 
labelled as AGB stars according to their position in the V-(U-V) CMD (see Section.~\ref{members}). 
Note that an incorrect attribution of a cluster star 
to an evolutionary sequence has a negligible impact on the derived abundances: 
a difference of 0.1 $M_{\odot}$ in stellar mass leads to a variation in log~g of $\sim$0.06, 
corresponding to a variation in the [Fe/H] abundances of 0.01 dex or less.

Due to the small ($\sim$15) number of Fe~I lines available for the GIRAFFE 
targets (that are the majority of the sample), microturbulent velocities ($v_t$) 
derived spectroscopically could be highly uncertain and we prefer to adopt the log~g--$v_t$ calibration 
provided by \citet{kirby09}. 
Although the small number of lines, we checked for most of the stars that 
the adopted $v_t$ does not provide significant trend between iron abundance and line strength. 
Only for a few stars we find a significant slope (at a level of 3$\sigma$ or more) 
and we change the value of $v_t$ in order to erase this trend.

The six stars observed with UVES allow to perform some sanity checks 
on the atmospheric parameters adopted for the analysis. 
 The atmospheric parameters have been derived spectroscopically according 
to three criteria: (i)~no trend between Fe~I abundance and excitation potential 
(to constrain $T_{\rm eff}$ ); (ii)~no trend between Fe~I abundance and line strength 
(to constrain  $v_t$); (iii)~same abundance from Fe~I and Fe~II lines 
(to constrain log~g).  
The spectroscopic $T_{\rm eff}$ well agree with the photometric ones, with an average 
difference (spectroscopic minus photometric) of --47 K ($\sigma$=~29 K), as well as 
the microturbulent velocities (the spectroscopic ones are on average lower 
than those obtained with the \citet{kirby09} calibration by --0.2 \kms, 
$\sigma$=~0.07 \kms). Finally, the spectroscopic gravities are lower by --0.2 dex 
($\sigma$=~0.03 dex), due to the small difference ($\sim$--0.1 dex) 
between [FeI/H] and [FeII/H] obtained with the photometric log~g.
The average difference between [FeI/H] derived from spectroscopic  
and photometric parameters is --0.03 dex ($\sigma$=~0.04 dex), 
suggesting that the adopted parameters are not affected by any 
significant bias.

\section{Chemical analysis}

Abundances of Fe, Al and Mg have been derived for all the targets, while 
abundances of O and Na have been determined only for the UVES targets, because 
no O and Na lines are available in the GIRAFFE HR21 setup. 
In particular, Fe~I lines are available for both UVES and GIRAFFE targets, 
while Fe~II lines have been measured only in UVES spectra 
(because no Fe~II lines are in the HR21 GIRAFFE grating).

For all these elements, the chemical analysis is based on 
1-dimensional, plane-parallel model atmospheres
calculated with the code {\tt ATLAS9} \citep{castelli05}, adopting 
$\alpha$-enhanced chemical mixture and without the inclusion of the 
approximate overshooting in the calculation of the convective flux. 
The derived abundance ratios are referred to the solar abundances by \citet{gs98} but 
for O for which we adopted the solar value by \citet{caffau11}.

The abundances of Fe, Na and Al have been calculated from the measured 
equivalent widths of metallic lines using the code {\tt GALA} \citep{mu13g}, 
based on the {\tt WIDTH9} software originally developed by R. L. Kurucz 
\citep[see][for details]{castelli05}. 
We selected transitions predicted to be unblended at the resolution of 
UVES and HR21 GIRAFFE setup. The atomic data are from the Kurucz/Castelli 
linelist\footnote{http://www.oact.inaf.it/castelli/castelli/linelists.html}, 
improved with new atomic data for some lines of interest 
(we refer the reader to \citet{mu17b} for a detailed 
description of the linelist). Table~\ref{linelist} lists the atomic data 
(wavelength, oscillator strength and excitation potential) for all the 
used transitions.

Al abundances from GIRAFFE targets are derived from the doublet at 8772-8773 \AA\ ,  
while for the UVES targets from the doublet at 6696-6698 \AA\ .
For some stars these lines are too weak to be detected, due to the low Al abundances 
(see Section~\ref{mgalab}), and only upper limits can be provided by adopting 
the abundance corresponding to an EW equal to 3 times the uncertainty calculated 
according to the \citet{cayrel88} formula.

The Na abundances for the UVES targets have been obtained from the doublets 
5682-5688 \AA\ and 6154-6160 \AA\ and corrected for departures from 
local thermodynamic equilibrium according to \citet{gratton99}.

The abundances of Mg for the GIRAFFE targets have been derived from the Mg~I line at 8806 \AA\ 
by using spectral synthesis, in order to properly account for the 
profile of this strong line that can have significant wings.
In particular, the abundances have been obtained using our own 
code {\tt SALVADOR} that performs a $\chi^2$-minimization between 
the observed spectra and a grid of synthetic spectra calculated 
on the fly with the spectral synthesis code {\tt SYNTHE}.
For the UVES targets, Mg abundances have been obtained from the measurement 
of the EW of the Mg~I line at 5711 \AA\ .
The Mg abundance for the GIRAFFE targets have been decreased by --0.28 dex 
in order to match the peak of the [Mg/Fe] distribution for the GIRAFFE stars 
with the highest [Mg/Fe] derived for the UVES targets. Offsets between the Mg abundances 
derived from the line at 8806 \AA\ and the optical lines have been already 
found in other analysis \citep[see e.g.][]{mu17}.
Note that we are interested mainly to the star-to-star variations of [Mg/Fe] in this cluster 
and not to the absolute value of the abundances.

The uncertainties in any abundance ratio have been calculated by summing in quadrature 
the errors arising from the measure procedure (EWs or spectral fitting) and from the 
atmospheric parameters.\\ 
{\sl (1)~Uncertainties due to EWs---} 
The errors in abundance due to the EW measurements have been computed as 
the dispersion of the mean normalized to the root mean square of the number of used lines. 
When only one line is available, the uncertainty is estimated considering the 
error in EW provided by {\tt DAOSPEC}.\\ 
{\sl (2)~Uncertainties due to spectral fitting---}
The uncertainties in the fitting procedure have been estimated 
using Montecarlo simulations. For different values of SNR, corresponding to the range 
of SNR covered by the GIRAFFE targets, samples of 500 synthetic spectra at the same spectral resolution and 
pixel-size of the GIRAFFE spectra and with the inclusion 
of Poissonian noise have been created and analyzed with the same fitting procedure 
used for the observed spectra. The dispersion of the derived abundance distribution is 
assumed as 1$\sigma$ uncertainty associated to the abundance from spectra with that SNR. 
The derived uncertainties in the fitting procedure range from $\sim$0.01 dex for SNR=260 to 
$\sim$0.07 dex for SNR=70 for the Mg~I line at 8806 \AA\ measured in the GIRAFFE targets. 
For the O abundance the typical uncertainty is about 0.05 dex.\\ 
{\sl (3)~Uncertainties due to the atmospheric parameters---}
These uncertainties are computed by varying each time only one parameter by the corresponding 
error, keeping the other ones fixed and repeating the analysis. 
According to the photometric uncertainties we estimated internal errors in $T_{\rm eff}$ 
of about 40 K, in log~g of 0.1 and in $v_{\rm t}$ of 0.1 \kms. Note that we are mainly 
interested to the internal uncertainties in the atmospheric parameters because the main goal 
of this study is investigate possible intrinsic star-to-star scatters.

\section{Membership}
\label{members}
\begin{figure*}
\plotone{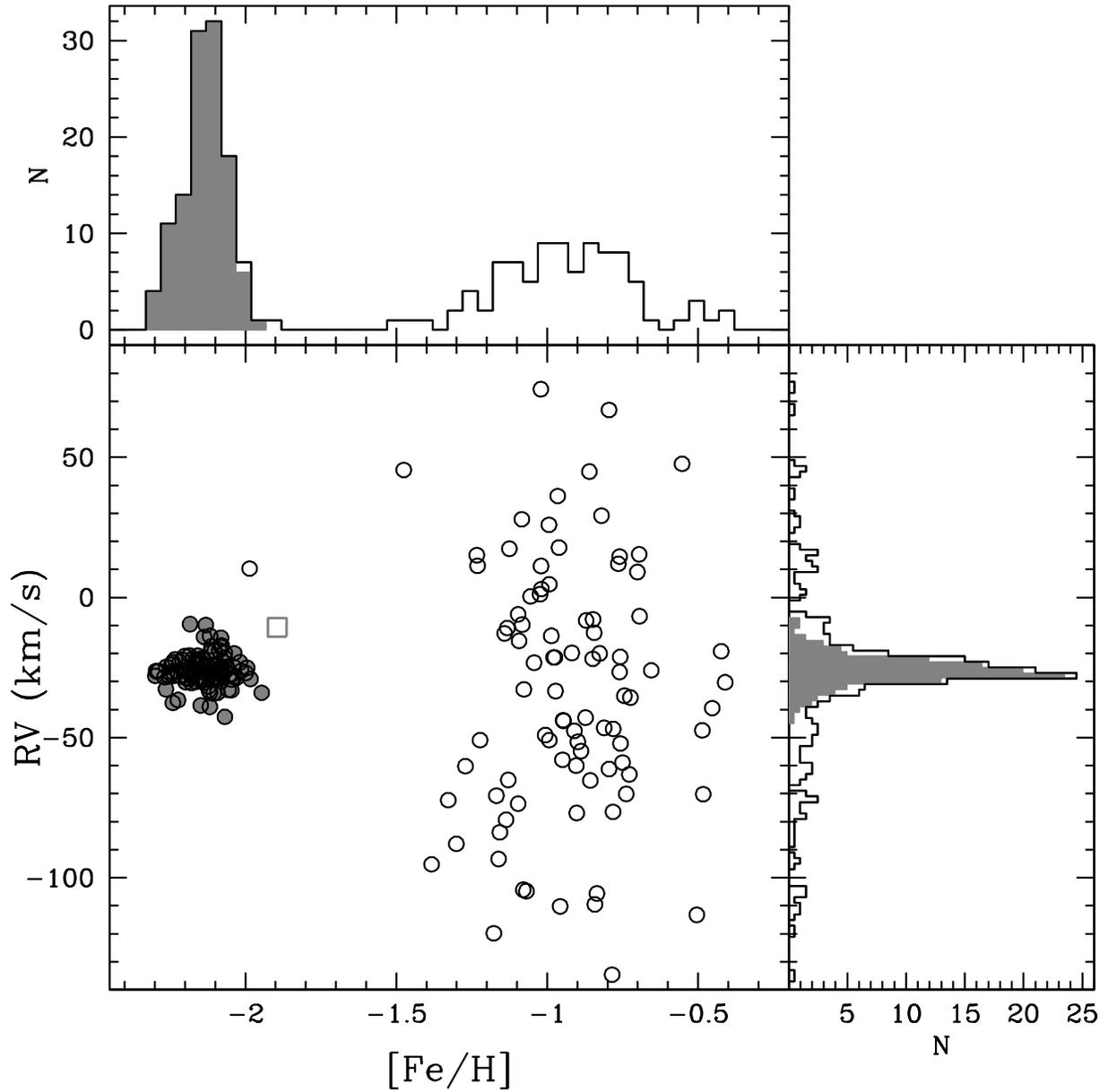}
\caption{The main panel shows the behaviour of the RV of the observed stars as a function 
of the [Fe/H] (same symbols of Fig.~\ref{rvd}). The histograms of [Fe/H] and RV distributions 
are also plotted (the grey shaded histograms include only cluster member stars).}
\label{rvfe}
\end{figure*}

We identified likely cluster member stars according to their RV and [Fe/H].
Fig.~\ref{rvfe} shows the position of the observed targets in the RV-[Fe/H] plane.
Stars belonging to NGC~5824 are easily identifiable as a clump of stars 
around RV$\sim$--25 \kms and [Fe/H]$\sim$--2.1 dex.
The surrounding field stars show a large RV distribution peaked at values similar 
to those of the cluster stars but metallicities larger than 
[Fe/H]$\sim$--1.5 dex and peaked at $\sim$--0.9 dex. 
We consider member stars those with [Fe/H] between --2.35 and --1.90 dex 
and with RV between -50 and 0 \kms (and marked as grey circles in Fig.~\ref{rvfe}).

\begin{figure*}
\plotone{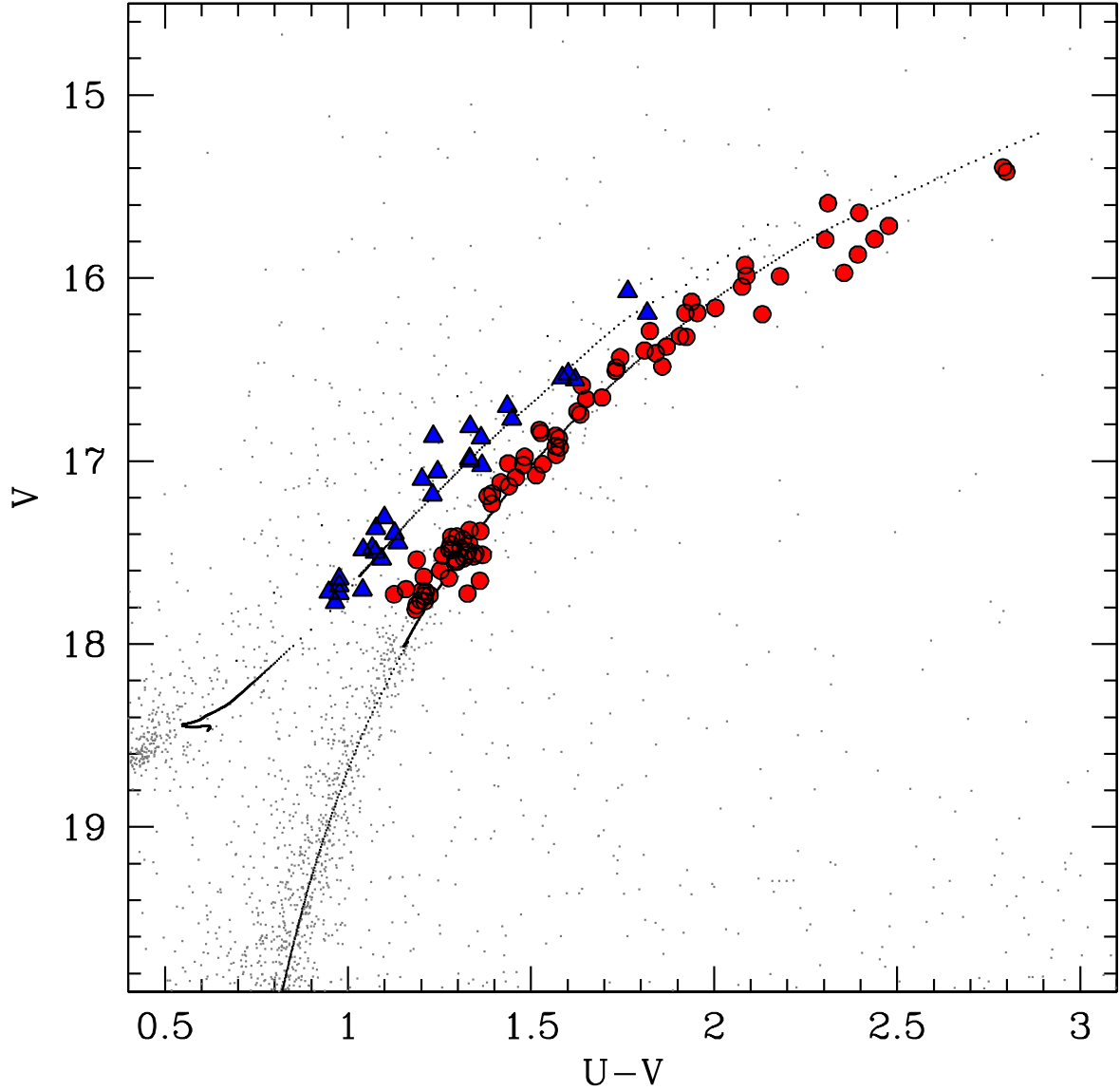}
\caption{Position of the member stars in the V-(U-V) CMD 
\citep{sanna14}: RGB and AGB stars are plotted as red circles and blue triangles, respectively. 
The best-fit theoretical isochrone from the BaSTI database 
\citep[13 Gyr, Z=~0.0006, $\alpha$-enhanced chemical mixture,][]{pietr06} is plotted as reference.}
\label{cmd1}
\end{figure*}

A total of 117 member cluster stars are identified. 
All the main information (coordinates, magnitudes, RVs) of these stars are listed 
in Table~\ref{info1}.
Note that 11 of them are located at a distance from the cluster center larger than 350 arcsec, 
where the sample is dominated by Galactic field stars. However, their metallicity 
is clearly different with respect to that of the surrounding field and this provides
their membership to the cluster. 
The mean heliocentric RV of this sample is -26.0$\pm$0.5 \kms ($\sigma$=~5.4 \kms), 
in good agreement with those quoted by DC14 e Ro16.
According to the distance in the V-(U-V) CMD 
to the best-fit theoretical isochrone \citep[see][for details]{sanna14},
we attributed each cluster star to RGB or AGB,
identifying 87 RGB stars and 30 AGB stars.
The position of the member stars on the V-(U-V) CMD 
(and their attribution to RGB or AGB sequences) is shown in Fig.~\ref{cmd1}. 
Note that attribution to a given evolutionary sequence of the bluest stars brighter than V$\sim$16.6 
is not trivial. Five stars with V$<$16.6 are labelled as AGB stars but we cannot exclude 
that they are RGB stars.

\section{The iron spread}
\label{iron}

We used the maximum likelihood (ML) algorithm described in \citet{mu12} 
to estimate whether the observed scatter measured in the iron abundance is compatible or not 
with a null intrinsic spread, taking into account the uncertainties 
of individual stars. 
Iron abundances and corresponding uncertainties are listed in Table~\ref{info2} and \ref{info3} 
for UVES and GIRAFFE targets respectively.
We checked different sub-samples of member stars.
\begin{enumerate}
\item Total sample (117 stars): 
the ML algorithm provides an average value of [Fe/H]=--2.14$\pm$0.01 dex
with an intrinsic spread of 0.02$\pm$0.01 dex (the observed scatter is 0.07 dex). 
This result suggests the presence of a small abundance spread.
\item RGB stars (87 stars): 
the average [Fe/H] of the RGB stars only is --2.11$\pm$0.01 dex with an 
observed scatter of 0.06 dex and an intrinsic spread of 0.00$\pm$0.01 dex.
\item AGB stars (30 stars): 
the average [Fe/H] of the AGB stars only is --2.20$\pm$0.01 dex with an 
intrinsic spread of 0.00$\pm$0.01 dex and an observed scatter of 0.07 dex.
If we exclude from the AGB sample the 5 brightest AGB stars with doubtful 
attribution the derived abundance and intrinsic scatter do not significantly change.
\item RGB stars in common with DC14 (66 stars): 66 out 88 stars in common 
with DC14 belong to RGB and they provide [Fe/H]=--2.12$\pm$0.01 dex with an 
intrinsic scatter of 0.00$\pm$0.02 dex. 
\end{enumerate}

\begin{figure*}
\epsscale{0.7}
\plotone{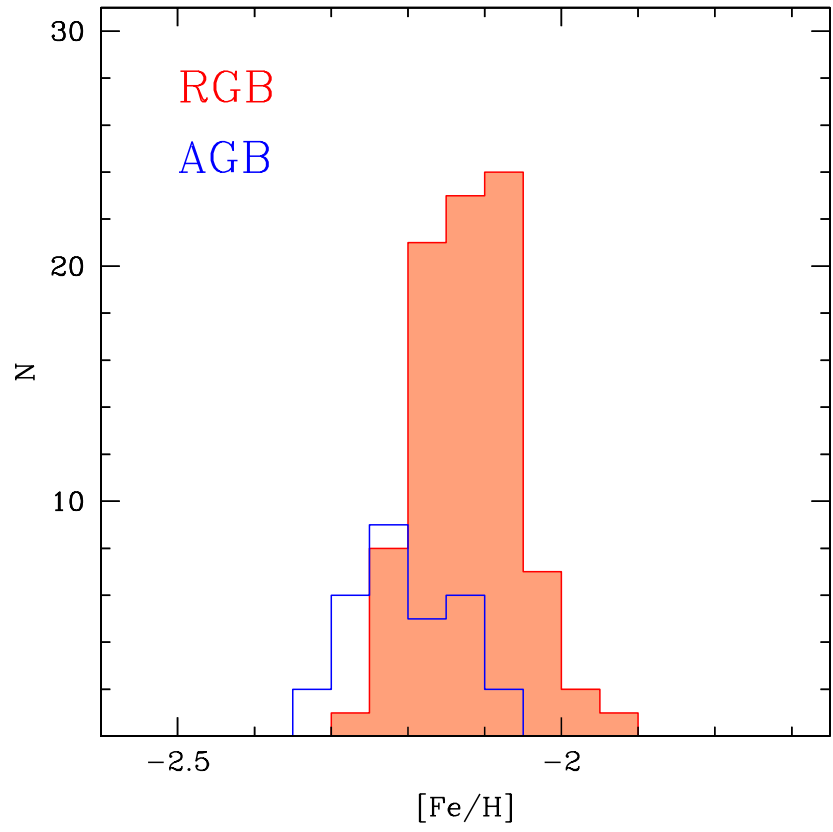}
\caption{[Fe/H] distributions derived from this study for RGB and AGB stars, 
red and blue histograms respectively.}
\label{ironh}
\end{figure*}

Even if the total sample seems to suggest a small (but marginally significant) iron dispersion 
(0.02$\pm$0.01 dex), when the sample of RGB and AGB stars are analyzed separately they provide 
a clear lack of intrinsic scatter pointing out that the cluster has an homogeneous 
iron content (as already pointed out by Ro16). 
AGB and RGB stars show a systematic difference in their [Fe/H] abundance ratios of 
about 0.1 dex, with the AGB stars having a lower abundance with respect to the RGB stars. 
Upper panel of Fig.~\ref{ironh} shows the [Fe/H] distribution derived for the 
RGB and AGB samples, red and blue histograms respectively, where the systematic 
offset between the two distributions is clearly visible.
This finding agrees with the chemical analyses of AGB stars in other GCs 
\citep{lp14,lp15,mu15,lp16}
that show a systematic (but still unexplained) underestimate of the [Fe/H] in AGB stars with 
respect to the RGB stars of the cluster when the neutral iron lines are used.
This difference between the iron abundances of RGB and AGB stars 
explains why the entire sample of cluster 
member stars provides a small iron dispersion (that totally disappears when 
the two groups of stars are analyzed independently).

\section{Light elements abundances: O, Na, Mg, Al}
\label{mgalab}

\begin{figure*}
\epsscale{1.0}
\plotone{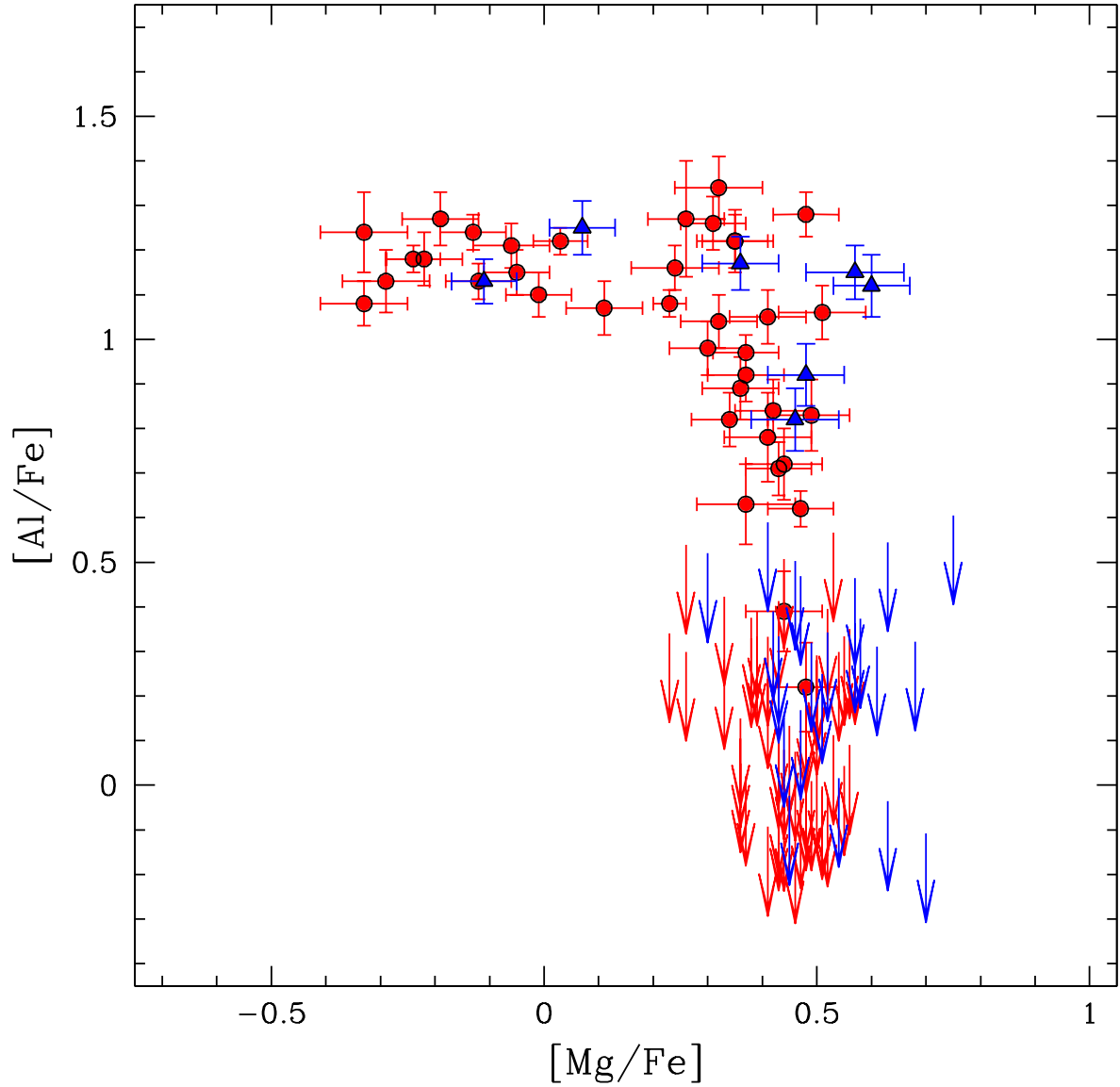}
\caption{Behaviour of [Al/Fe] as a function of [Mg/Fe] for the member stars 
of NGC~5824. Arrows indicate upper limits for [Al/Fe]. Red and blue symbols are for the RGB and AGB stars, 
respectively.}
\label{mgal}
\end{figure*}

We investigated the occurrence in NGC~5824 of chemical anomalies in the 
abundances of O, Na, Mg and Al. 
As suggested by Ro16, NGC~5824 has a star-to-star dispersion in the [Mg/Fe] 
abundance larger than those usually observed in most of the GCs 
\citep[see e.g][]{carretta09u,meszaros15,pancino17}. 
Fig.~\ref{mgal} shows the behavior of [Al/Fe] as a function of [Mg/Fe] as obtained from 
our study. 
Mg abundances are available for all the member stars and true measures 
of Al have been obtained only for 45 stars, while for the other targets 
upper limits are provided, due to the weakness of the Al lines. 
A large spread is observed both in [Mg/Fe] and [Al/Fe], 
with [Mg/Fe] ranging from enhanced values
down to sub-solar values (the minimum abundance is $\sim$--0.35 dex). 
On the other hand, [Al/Fe] ranges over more than 1 dex. 
A clear Mg-Al anticorrelation is observed, 
with all the stars with [Mg/Fe]$<$+0.15 dex having [Al/Fe]$\sim$+1.2 dex.

\begin{figure*}
\plotone{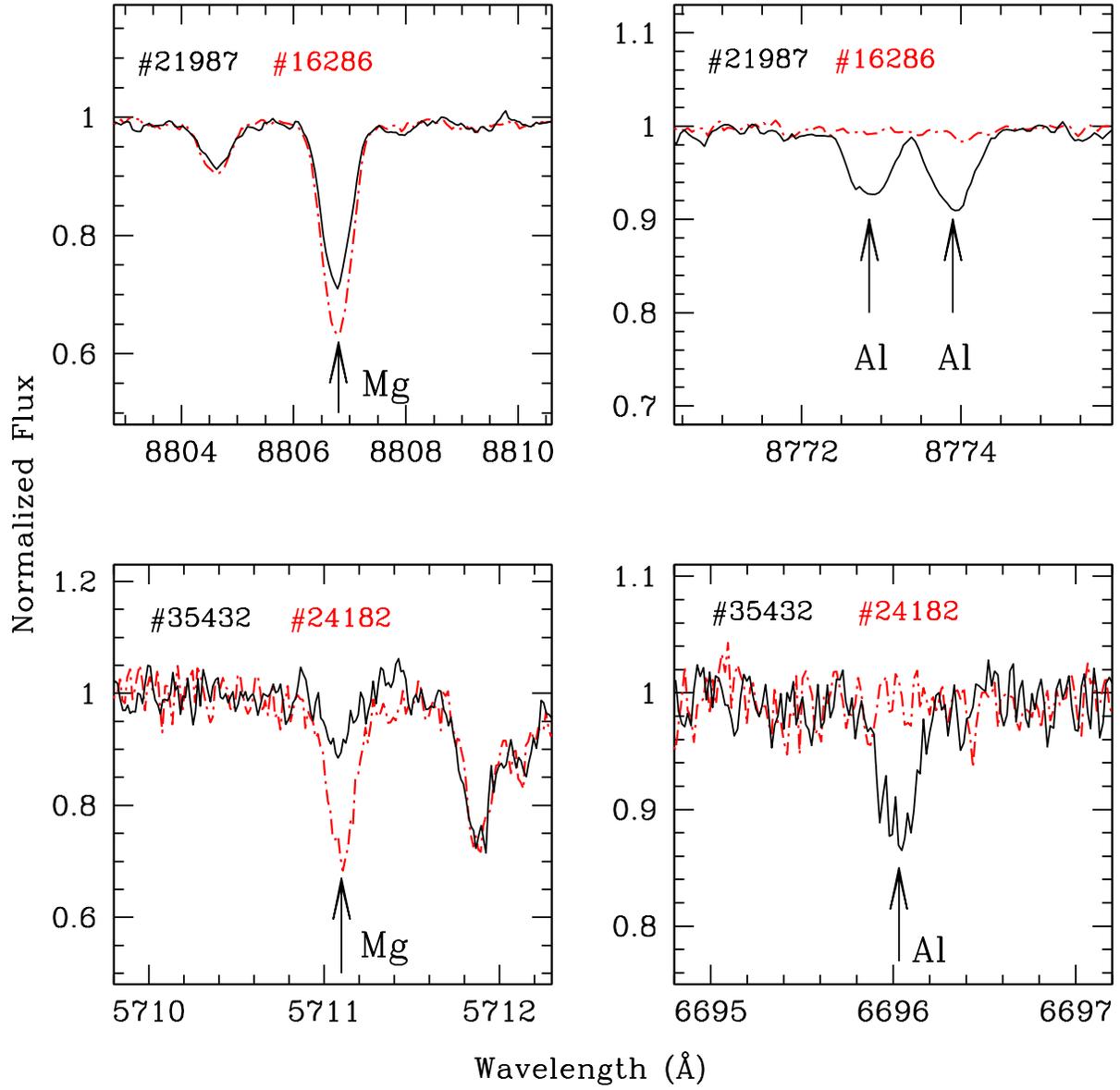}
\caption{Portions of the spectra of the GIRAFFE stars \#16286 and \#21987 (upper panels),
and of the UVES stars \#35432 and \#24162 (lower panels), around Mg and Al lines. 
The stars of both pairs have very similar atmospheric parameters.}
\label{mgalspec}
\end{figure*}

A direct evidence of this strong anticorrelation is visible in Fig.~\ref{mgalspec} 
where the spectra of two stars observed with GIRAFFE (namely \#16286 and \#21987) 
and of two stars observed with UVES (namely \#35432 and \#24182)
are compared around the Mg and Al lines. 
The stars of each pair have been selected in order to have very similar atmospheric 
parameters.
The star \#16286 shows a prominent Mg line but a total 
lack of the Al lines, while for the star \#21987 the situation is the opposite, 
with a weaker Mg line with respect the other star but well visible Al lines. 
The same situation is observed for the two UVES targets but using different 
Mg and Al features.
Being the atmospheric parameters of the stars in each pair very similar, the observed 
different strengths in Mg and Al lines can be ascribed only to differences in 
abundance.

\begin{figure*}
\plotone{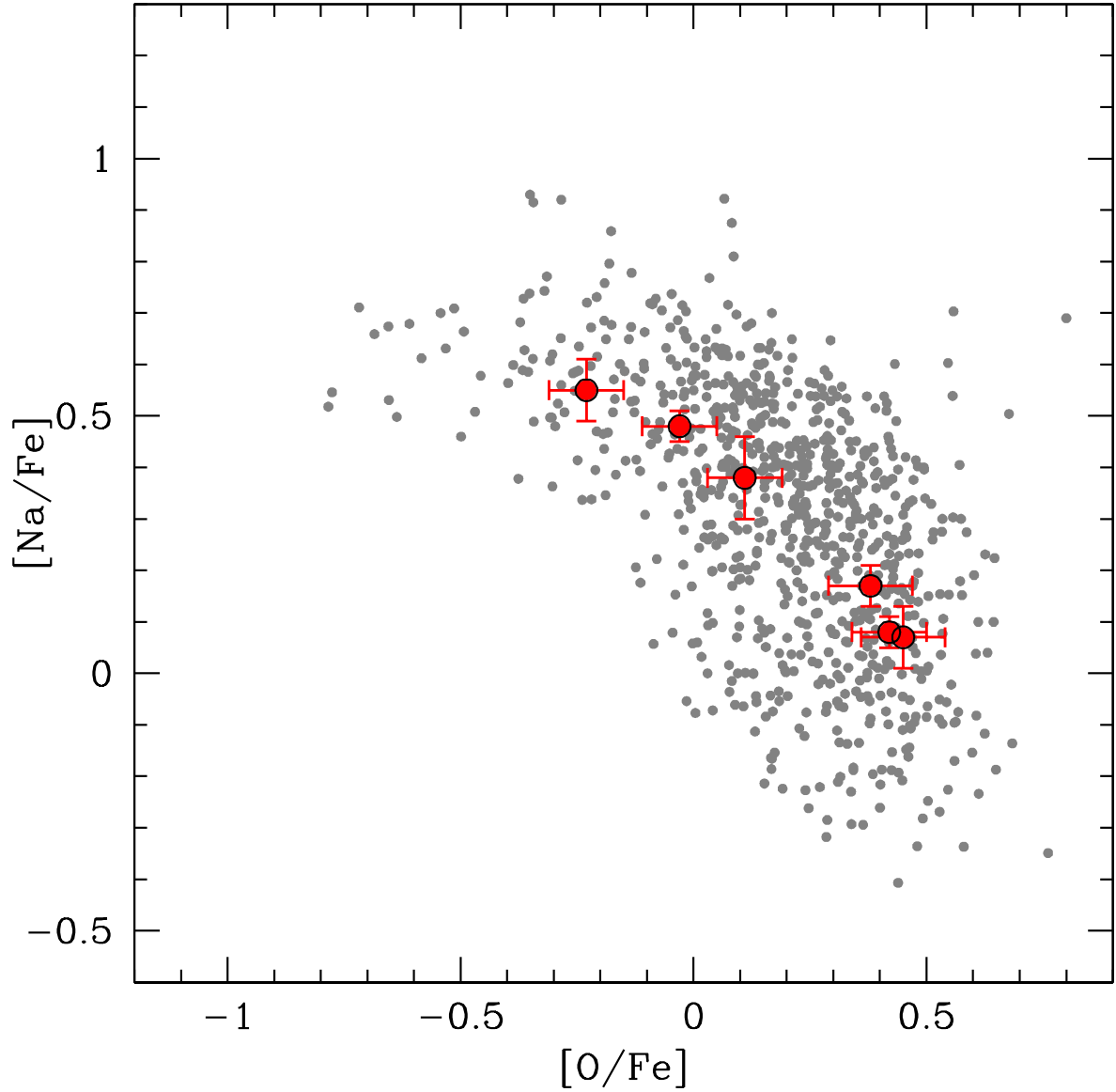}
\caption{Behaviour of [Na/Fe] as a function of [O/Fe] for the six RGB stars of NGC~5824 
observed with UVES (red circles) in comparison with the Galactic GC stars (grey circles) analyzed by 
\citet{carretta09g} and \citet{carretta09u}.}
\label{nao}
\end{figure*}

For the UVES targets we can measure also O and Na abundances. 
Fig.~\ref{nao} shows the behavior of the [Na/Fe] abundance ratio for 
the six RGB stars observed with UVES as a function of [O/Fe] (red points) 
in comparison with the stars measured in 19 Galactic GCs by \citet{carretta09g,carretta09u}. 
The six stars show a clear anti-correlation that well matches with that observed in 
other Galactic GCs.

\section{Comparison with Da Costa et al.(2014)}
\label{mcat}

The adopted GIRAFFE set-up allows to measure also the Ca~II triplet lines, 
used by DC14 to infer an indirect estimate of the iron content of the cluster.
Similar to what they did, we measured the EWs of the two strongest 
Ca~II lines (8542 \AA\ and 8662 \AA\ ) for the cluster member stars, adopting a Voigt profile 
in order to reproduce the damped wings of these lines.
[Fe/H] have been derived using the calibration provided by \citet{saviane12} and adopting a 
reference horizontal branch magnitude $V_{\rm HB}$=~18.50 as quoted by DC14. 
For the stars in common with DC14 we find a mean difference in the sum of the 
EWs of the two Ca~II lines of +0.03$\pm$0.01 \AA\  ($\sigma$=~0.13 \AA\ ). 
Such a small offset rules out the existence of any significant difference 
between the two studies, since a variation of 0.03 \AA\ in the summed EW translates in a difference 
in [Fe/H] of about 0.01 dex.

The uncertainties in the [Fe/H] derived from Ca~II triplet lines have been estimated 
considering three sources of errors: 
{\sl (i)}~the uncertainty in the fitting procedure of the Ca~II lines, 
{\sl (ii)}~that in the continuum location, and 
{\sl (iii)}~that related to the adopted EW-[Fe/H] calibration.
The first two sources have been estimated using Montecarlo simulations and 
analyzing samples of 500 synthetic spectra each with different SNR, and adopting the spectral resolution 
and pixel-size of the HR21 GIRAFFE setup. 
The uncertainty in the EW-[Fe/H] calibration by \citet{saviane12} has been estimated 
considering the global residual of their fit ($\sim$0.13 dex) normalized to the 
root mean square of the number of GCs used to derive the relation. 
The total uncertainties range from $\sim$0.09--0.10 dex for the faintest targets 
to $\sim$0.04 dex for the brightest ones.

The average [Fe/H] abundance derived from the 81 RGB stars observed with GIRAFFE and for which 
we measured the Ca~II lines strength is --1.99$\pm$0.01 dex with an intrinsic scatter 
of 0.02$\pm$0.01 dex, as derived adopting the ML approach.
When we consider only the RGB stars in common with DC14 (60 in total) 
the average abundance is [Fe/H]=--1.99$\pm$0.01 dex with an intrinsic scatter 
of 0.00$\pm$0.01 dex.

\begin{figure*}
\plotone{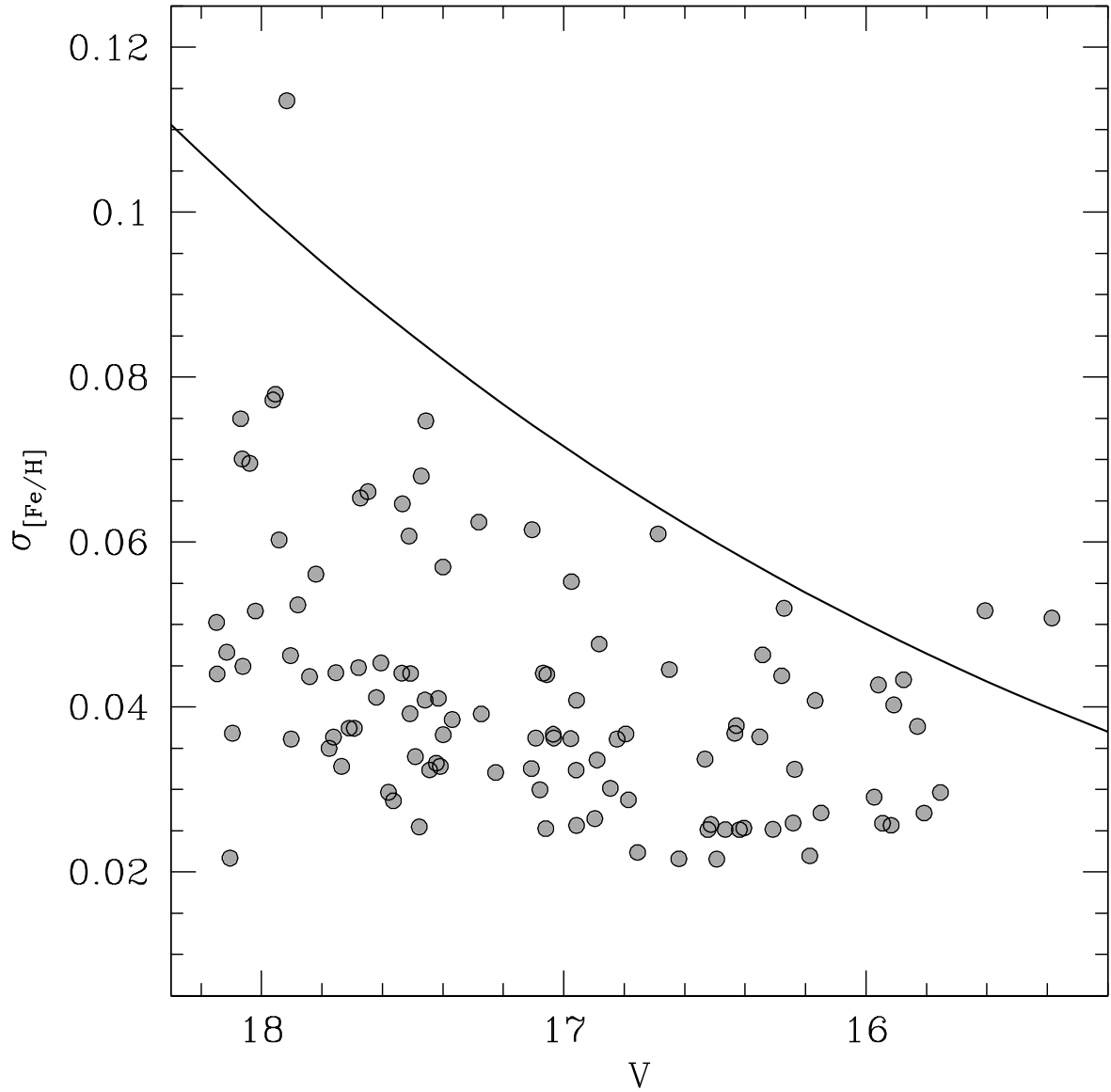}
\caption{Behaviour of [Fe/H] uncertainties from DC14 as a function of the V-band magnitude 
(grey points) in comparison with the expected uncertainty according to our Montecarlo simulations 
(black line).}
\label{errfe}
\end{figure*}

When the iron abundances of the entire sample of 108 RGB stars studied by DC14 is analyzed 
with the ML approach,  an average abundance of --2.01$\pm$0.01 dex is found, 
with an intrinsic spread of 0.05$\pm$0.01 dex, suggesting a small but not 
negligible iron spread, at variance with the result obtained with our dataset. 
In this test the uncertainties in EWs quoted by DC14 
have been transformed into [Fe/H] errors according to the relation by \citet{saviane12}.
The uncertainties in their [Fe/H] have been re-derived using the same procedure described 
above: samples of synthetic spectra with different SNR between 50 and 130 
(see Fig.~11 in DC14) have been simulated, including a spectral resolution of $\sim$2500 and
a pixel-size of 0.8 \AA\ per pixel, in order to simulate spectra similar to those obtained with FORS2 
by DC14. 
Fig.~\ref{errfe} shows the behavior of the abundance errors (derived from the 
uncertainties in EWs quoted by DC14) as a function of the V-band magnitude for 
the DC14 targets (grey circles). In comparison, the black line represents the 
uncertainties expected according to our Montecarlo simulations. 
On average the Montecarlo uncertainties are systematically higher than those quoted 
by DC14. 

When the new set of errors is adopted, the ML algorithm provides an intrinsic spread 
of 0.01$\pm$0.02 dex, 
similar to what we obtained with the [Fe/H] derived from the Ca~II triplet measured 
with GIRAFFE spectra, but in contrast with the result obtained with the uncertainties 
by DC14. This finding suggests that the small iron spread obtained by DC14 was probably due to an 
underestimate of the uncertainties in the Ca~II lines EWs.
It is worth noting that the observed scatter of the [Fe/H] distribution 
of the DC14 study is quite small and comparable with the scatters usually observed 
in GCs studied with high spectral resolution 
\citep[with scatter smaller than 0.05-0.06 dex, see][]{carretta09}.

Because of the large spread in [Mg/Fe] found in NGC~5824 and not observed in most of the globulars, 
we check for possible correlations between the iron abundance derived from 
the Ca~II lines and [Mg/Fe]. In fact, Mg plays a relevant role in the opacity of 
giant star atmospheres, being one of the most important electron donors. 
A significant depletion in the Mg abundances leads to a decrease of the $H^{-}$ opacity 
and of the electronic pressure, thus increasing the 
line strength of the Ca~II lines (at a constant Fe and Ca abundance). 
This effect has been revealed for the first time in the GC NGC~2419 \citep{mu12} 
where the unusual Mg depletion (down to [Mg/Fe]$\sim$--1 dex) observed in some 
cluster stars leads to a significant increase of the EWs of the Ca~II lines 
and that can be erroneously interpreted as a high Fe abundance.

\begin{figure*}
\plotone{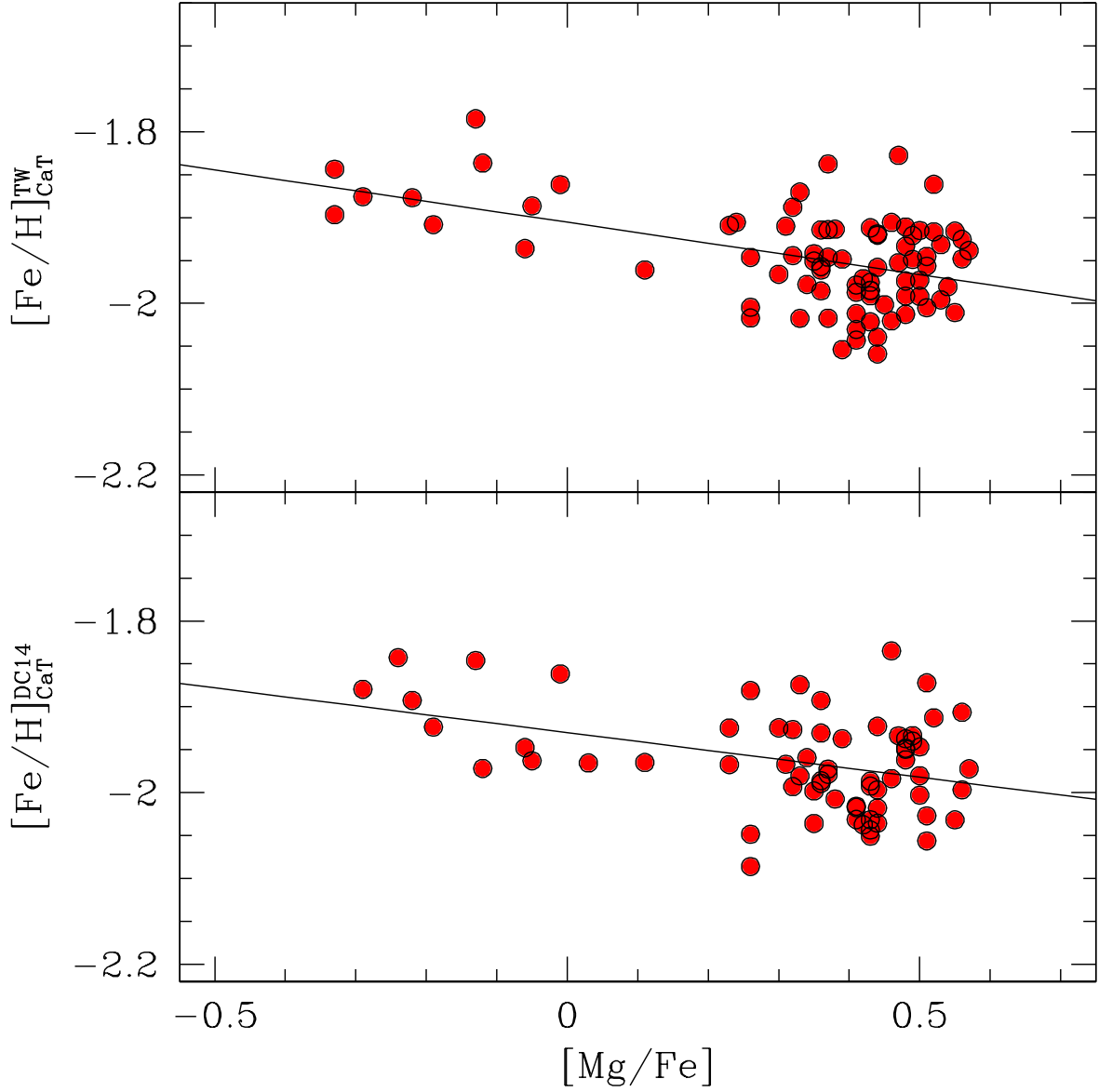}
\caption{Behaviour of [Fe/H] as derived from the Ca~II triplet lines (from this work, upper panel, and 
from DC14, lower panel) as a function of [Mg/Fe]. Solid grey lines are the linear best fits. }
\label{mgcat}
\end{figure*}

Fig.~\ref{mgcat} shows the behavior of [Fe/H] (as derived from Ca~II lines) 
as a function of [Mg/Fe] considering RGB stars only measured from our sample 
(upper panel) and from that by DC14 (lower panel). 
Both datasets show a mild anticorrelation between [Fe/H] and [Mg/Fe] in the 
expected sense: the stars with low [Mg/Fe] have systematically higher [Fe/H] 
abundances (solid lines in Fig.~\ref{mgcat} show the best linear fit to the data).
The Spearman correlation test provides probabilities that the variables 
are non-correlated of 0.01 from our sample and of 0.06 from the [Fe/H] by DC14. 
The probability of correlation increases when the [Fe/H] inferred from 
the GIRAFFE spectra are used, likely due to the higher spectral resolution and SNR 
with respect to the dataset by DC14 (that have SNR between 50 and 120, while the 
spectra of this study between 70 and 250).
This finding contradicts the analog test performed by Ro16 that found a low probability 
of correlation between their [Mg/Fe] and the [Fe/H] provided by DC14. 
Such a difference is likely due to the  higher uncertainty of the 
[Mg/Fe] derived by Ro16 (the typical SNR of their spectra is 40-50) and 
of the [Fe/H] by DC14 with respect to this study.

The above correlation is weak and the few Mg-poor stars do not significantly contribute to 
change the observed scatter: when the stars with sub-solar [Mg/Fe] are excluded, 
the dispersion in the average [Fe/H] decreases from 0.06 down to 0.05 dex with both 
the datasets. However, the detection of this 
anticorrelation (using two independent sets of Ca~II lines strengths) 
confirms that metallicities inferred from Ca~II lines could be over-estimated 
in Mg-poor stars and, in general, could be biased in case of anomalous and/or 
exotic chemical compositions (like in NGC~2419).

\section{Comparison with Roederer et al.(2016)}

23 stars of our spectroscopic sample are in common with that of Ro16. 
When we compare the atmospheric parameters of these targets we find 
a significant difference in the adopted $T_{\rm eff}$ scales, with a 
difference (our study - Ro16) of +139 K ($\sigma$=~45 K). 
For the gravity the average difference is +0.09 ($\sigma$=~0.03) 
and for $v_t$ is +0.00 \kms ($\sigma$=~0.26 \kms). 
Ro16 estimated the atmospheric parameters with an approach similar
to that followed in our analysis, deriving $T_{\rm eff}$ and log~g from the 
photometry, but constraining spectroscopically $v_t$ 
\citep[while we adopted the log~g-$v_{\rm t}$ calibration by][]{kirby09}
\footnote{Even if we adopted the same photometric magnitudes, E(B-V), exctintion 
coefficients and transformation between photometric systems adopted by Ro16, 
we are not able to reproduce the $T_{\rm eff}$ quoted in their Table 1. 
We note that the only way to recover the same values is to assume a E(B-V)=~0.07 mag 
instead of the quoted value of E(B-V)=~0.14 mag.}.
The slightly lower log~g found by Ro16 are compatible with the lower $T_{\rm eff}$ that they adopted.
This difference in $T_{\rm eff}$ explains also the difference in the derived Fe abundances, 
with an average difference of +0.23 dex ($\sigma$=~0.07 dex).
In their discussion of the metallicity of NGC~5824, Ro16 used [FeII/H] because they found 
a large ($\sim$0.4 dex) difference between Fe abundances from neutral and single ionized lines, 
with [FeI/H] systematically lower than [FeII/H]. 
As a check, we re-analysed the stars in common with Ro16 adopting their atmospheric 
parameters and finding [FeI/H]=--2.29$\pm$0.02 dex ($\sigma$=~0.08 dex) 
and [FeII/H]=--2.07$\pm$0.01 dex ($\sigma$=0.02 dex, only for the 6 UVES stars). 
They attributed the observed difference between [FeI/H] and [FeII/H] to the 
fact that FeI lines can be less reliable diagnostics with respect to FeII lines 
in metal-poor giant stars (for instance due to the occurrence of non-local 
thermodynamical equilibrium). We think that their measured difference between 
the two abundance ratios is instead due to an underestimate of the adopted temperatures.
This is confirmed by the re-analysis of the UVES targets using the Ro16 $T_{\rm eff}$, 
that providing significant positive slopes between 
the iron abundance and the excitation potential, pointing out an under-estimate of $T_{\rm eff}$. 
On the other hand, our photometric $T_{\rm eff}$ for the UVES stars do not require 
adjustment to reproduce the excitation equilibrium. 

The derived [Mg/Fe] distribution is compatible with that obtained by Ro16.
Ro16 found a large difference ($\sim$0.4 dex) between the Mg abundances 
derived from two stars observed both with M2FS (using the Mg~I line at 4571.1 \AA\ ) 
and MIKE (where abundances are based on three optical lines), probably 
due to departures from local thermodynamical equilibrium that could affect that 
blue transition at 4571.1 \AA\ arising from 
the ground level of the Mg~I atom. Following the suggestion by Ro16, we increase by 0.4 dex 
all their abundances derived from M2FS spectra, finding an average difference for the stars 
in common between our and their study of +0.10$\pm$0.03 dex ($\sigma$=~0.15 dex). 
Despite of possible offsets between the two abundance scales, the
range of [Mg/Fe] in the two studies, based on different Mg transitions, is fully compatible, 
confirming the large depletion of [Mg/Fe] in some stars of NGC~5824.

\section{Conclusions}
From the analysis of FLAMES high-resolution spectra of 117 giant stars members of 
the GC NGC~5824 we obtained two main results, namely 
(1)~the lack of iron abundance spread, and (2)~the detection of an extended Mg-Al anticorrelation.

The cluster has an average iron abundance of --2.11$\pm$0.01 dex when we consider the sample 
of 87 {\sl bona-fide} RGB stars.
This sample does not show any evidence of intrinsic spread, confirming the first claim 
by Ro16 (that analyzed 26 stars with V$<$17) but based on a larger sample. 
According to this finding, NGC~5824 turns out to be a {\sl normal} GC, 
not showing any evidence of internal self-enrichment in terms of iron, hence there is no 
reason to consider it the remnant of a complex stellar system (like a dwarf 
spheroidal galaxy). On the other hand it does not show any significant evidence 
of chemical peculiarities (as spread in s-process or C+N+O elements). 
In fact, although neither our dataset not that by Ro16 allowed to directly measure C and 
N abundances, the optical CMDs of NGC~5824 \citep[see][]{sanna14,walker17} did not 
reveal any anomalous splitting of the sub-giant branch 
\citep[which has been interpreted as an evidence of an intrinsic C+N+O spread, see][]{piotto12}.
Also, as shown by Ro16, no star-to-star variation in s-process elements 
is found in NGC~5824, with the only exception of one star in the Ro16's sample that 
exhibits a systematic enhancement of the s-process elements. This star is present 
also in our sample (the UVES target \#24182): we found the same enhancement of 
the s-process lines with respect to the other five UVES targets (no s-process transitions 
are available in the GIRAFFE spectra). However, this star shows an iron abundance 
fully compatible with the other stars, in contrast with the other anomalous 
GCs where the most metal-rich stars exhibit higher s-process abundances.

The present sample allows also to detect the occurrence of the standard signature 
of multiple populations in GCs, the Na-O anticorrelation. 
Additionally, we detect a very extended Mg-Al anticorrelation. 
At variance with the Na-O anticorrelation that is observed in all the old GCs 
in our Galaxy and in other galaxies of the Local Group, the Mg-Al anticorrelation 
turns out to be present only in a few clusters.
This is likely due to the different temperatures needed to ignite the 
NeNa cycle (responsible for the Na-O anticorrelation, T$\sim$50 MK) 
and the MgAl cycle (responsible for the Mg-Al anticorrelation, T$\sim$70 MK).

First \citet{carretta09u} proposed that the spread in [Al/Fe] and the extension 
of the Mg-Al anticorrelation are driven by two main parameters, namely 
the present-day cluster mass and the metallicity. Subsequent analyses by 
\citet{meszaros15} and \citet{pancino17} confirmed this behavior. 
The spread in [Al/Fe] decreases increasing the metallicity, 
with the most metal-rich and low-mass GCs showing little or no [Al/Fe] spreads
\citep[see the cases of M4, \citet{carretta09u} and NGC~6362,][]{massari17}. 
Well-developed Mg-Al anticorrelations have been found in metal-poor GCs 
(NGC~4833, M15, M92) or in high-mass GCs with intermediate 
(NGC~6752, M13) or high metallicity (NGC~2808). 
\citet{pancino17} show that the spread in [Al/Mg] abundance ratio 
clearly increases with the present-day mass and decreases with metallicity.
Also, Mg-poor stars (with [Mg/Fe]$<$+0.1 dex) have been found in 
NGC~2419 \citep{mu12,cohen12}, M54 \citep{carretta10b} and in 
Omega Centauri \citep{norris95}, the most metal-poor, massive GC-like systems. 

NGC~5824 well fits into this framework, being the 14$^{th}$ most luminous GC 
\citep[according to ][]{harris10} and the third most luminous GC among 
those with [Fe/H]$<$--1.9 dex, after NGC~2419 and M15 
(both of them showing a large [Mg/Fe] spread but homogenous [Fe/H]). 

This behavior of the [Al/Fe] and [Mg/Fe] spreads with the metallicity is 
in principle compatible with the scenario where the main polluters 
are the high-mass AGB stars of the first GC stellar generation that can active the MgAl 
cycle during hot bottom burning. 
At low metallicity the AGB stars can active the hot bottom burning 
at lower masses \citep{ventura01, ventura13,ventura16}. 
On the other hand, massive GCs, even at high metallicity, have deeper 
potential wells and they can more efficiently retain the polluter ejecta 
(like in the case of NGC~2808).

We conclude that NGC~5824 is a standard globular cluster, without spread in [Fe/H] and 
with the presence of usual chemical anomalies (both Na-O and Mg-Al anticorrelations), but 
showing a large (and rare) spread in Mg.

\acknowledgments
We thank the referee, Luca Sbordone, for his useful suggestions and comments.

\begin{deluxetable}{lccccccc}
\tablecolumns{8} 
\tablewidth{0pc}  
\tablecaption{Main information on the spectroscopic targets of NGC~5824.}
\tablehead{ 
\colhead{ID} &   \colhead{${\rm ID_{DC14}}$} &  RA      &   Dec     & U & V &  ${\rm RV_{hel}}$  &  Sequence  \\
             &                       & (J2000)  &  (J2000)  &   &   &  (\kms)    &    }
\startdata 
\hline   
 8953	&  ---  	   &	226.1162748 &	--33.1975713 &  18.464  &  17.032  & --24.30$\pm$0.10   & ${\rm RGB}$  \\  
10222	& 62000309	   &	226.0997769 &	--33.1493884 &  18.503  &  17.530  & --22.10$\pm$0.35   & ${\rm AGB}$  \\  
10647	&  ---  	   &	226.0940622 &	--33.1212687 &  17.901  &  15.773  & --33.00$\pm$0.25   & ${\rm RGB}$  \\  
10967	&  ---  	   &	226.0904849 &	--32.9280129 &  18.218  &  15.972  & --26.70$\pm$0.06   & ${\rm RGB}$  \\  
11276	& 42007983	   &	226.0868524 &	--33.0699666 &  18.354  &  17.224  & --28.60$\pm$0.12   & ${\rm AGB}$  \\  
11730	&  ---  	   &	226.0817615 &	--32.9692462 &  18.085  &  16.306  & --30.30$\pm$0.06   & ${\rm RGB}$  \\  
12035	&  ---  	   &	226.0783242 &	--33.0665547 &  18.586  &  17.645  & --28.10$\pm$0.03   & ${\rm AGB}$  \\  
12077	&  ---  	   &	226.0779387 &	--33.0395730 &  18.033  &  16.367  & --26.30$\pm$0.03   & ${\rm AGB}$  \\  
12898	& 42011701	   &	226.0699481 &	--33.0487649 &  19.097  &  17.890  & --29.10$\pm$0.14   & ${\rm RGB}$  \\  
13068	& 42008963	   &	226.0685097 &	--33.0643136 &  18.780  &  17.703  & --23.80$\pm$0.11   & ${\rm RGB}$  \\  
13705	&  ---  	   &	226.0629148 &	--33.1187793 &  18.379  &  17.165  & --28.00$\pm$0.06   & ${\rm AGB}$  \\  
13793	&  ---  	   &	226.0623981 &	--32.9325047 &  18.411  &  16.916  & --30.40$\pm$0.04   & ${\rm RGB}$  \\  
13894	& 11001198	   &	226.0614866 &	--33.0429962 &  18.136  &  16.463  & --23.80$\pm$0.06   & ${\rm RGB}$  \\  
14000	& 62000027	   &	226.0608134 &	--33.1705065 &  18.363  &  16.658  & --26.80$\pm$0.05   & ${\rm RGB}$  \\  
\hline
\enddata 
\tablecomments{$~~~~~$Identification number (from \citet{sanna14} and from DC14), coordinates, magnitudes, atmospheric 
parameters, heliocentric radial velocities and evolutionary sequences for all the member stars (this table is 
available in its entirety in machine-readable form.)}
\label{info1}
\end{deluxetable}

\begin{deluxetable}{lcccc}
\tablecolumns{5} 
\tablewidth{0pc}  
\tablecaption{Wavelength, species, oscillator strength, excitation potential of the used 
transitions. The last column indicates whether the transition has been measured in UVES or GIRAFFE spectra.}
\tablehead{ 
\colhead{Wavelength} &  \colhead{species} &  \colhead{log~gf} &  $\chi$      &  Instrument     \\
 \colhead{\AA}       &          &                   &  (eV)        &    }
\startdata 
\hline   
 4859.741 &   ${\rm Fe I }$	 &  -0.764    &  2.875  &  ${\rm UVES}$  \\  
 4882.143 &   ${\rm Fe I }$	 &  -1.640    &  3.417  &  ${\rm UVES}$  \\  
 4907.732 &   ${\rm Fe I }$	 &  -1.840    &  3.430  &  ${\rm UVES}$  \\  
 4917.230 &   ${\rm Fe I }$	 &  -1.160    &  4.191  &  ${\rm UVES}$  \\  
 4924.770 &   ${\rm Fe I }$	 &  -2.114    &  2.279  &  ${\rm UVES}$  \\  
 4938.814 &   ${\rm Fe I }$	 &  -1.077    &  2.875  &  ${\rm UVES}$  \\  
 4950.106 &   ${\rm Fe I }$	 &  -1.670    &  3.417  &  ${\rm UVES}$  \\  
 4969.917 &   ${\rm Fe I }$	 &  -0.710    &  4.217  &  ${\rm UVES}$  \\  
 4985.253 &   ${\rm Fe I }$	 &  -0.560    &  3.929  &  ${\rm UVES}$  \\  
 4985.547 &   ${\rm Fe I }$	 &  -1.331    &  2.865  &  ${\rm UVES}$  \\  
 4993.358 &   ${\rm Fe II}$	 &  -3.620    &  2.807  &  ${\rm UVES}$  \\  
 5001.863 &   ${\rm Fe I }$	 &  -0.010    &  3.882  &  ${\rm UVES}$  \\  
 5002.793 &   ${\rm Fe I }$	 &  -1.530    &  3.396  &  ${\rm UVES}$  \\  
 5014.942 &   ${\rm Fe I }$	 &  -0.303    &  3.943  &  ${\rm UVES}$  \\  
\hline
\enddata 
\tablecomments{$~~~~~$This table is 
available in its entirety in machine-readable form.}
\label{linelist}
\end{deluxetable}

\begin{deluxetable}{lcccccccc}
\tablecolumns{9} 
\tablewidth{0pc}  
\tablecaption{Atmospheric parameters and abundance ratios for the UVES targets of  NGC~5824.}
\tablehead{ 
\colhead{ID} &  $T_{\rm eff}$ &  log~g      &   $v_{\rm t}$     & [Fe/H] & [O/Fe] &  [Na/Fe]  &  [Mg/Fe]  & [Al/Fe]\\
               &        (K)        &   &  (\kms)  & (dex)  &  (dex) &  (dex)   & (dex) & (dex) }
\startdata 
\hline   
 20836     &  4270  &  0.77   &  2.0  & --2.07$\pm$0.06   &  --1.99$\pm$0.08  & 0.08$\pm$0.03  &  0.48$\pm$0.08 & $<$0.21  \\  
 24182     &  4252  &  0.83   &  1.9  & --2.16$\pm$0.06   &  --1.93$\pm$0.09  & 0.09$\pm$0.04  &  0.50$\pm$0.02 & $<$0.28  \\  
 26034     &  4238  &  0.69   &  2.0  & --2.14$\pm$0.06   &  --2.00$\pm$0.08  & 0.08$\pm$0.08  &  0.23$\pm$0.03 & 1.08$\pm$0.03\\  
 27416     &  4344  &  0.90   &  1.9  & --2.14$\pm$0.06   &  --2.01$\pm$0.09  & 0.09$\pm$0.06  &  0.38$\pm$0.03 & $<$0.33  \\  
 31793     &  4313  &  0.88   &  1.9  & --2.15$\pm$0.06   &  --1.96$\pm$0.08  & 0.08$\pm$0.03  &  0.03$\pm$0.05 & 1.22$\pm$0.03\\  
 35432     &  4239  &  0.78   &  2.0  & --2.17$\pm$0.06   &  --1.95$\pm$0.08  & 0.08$\pm$0.06  &--0.24$\pm$0.05 & 1.18$\pm$0.03\\  
\hline
\enddata 
\tablecomments{$~~~~~$}
\label{info2}
\end{deluxetable}

\begin{deluxetable}{lcccccc}
\tablecolumns{7} 
\tablewidth{0pc}  
\tablecaption{Atmospheric parameters and abundance ratios for the GIRAFFE targets members of NGC~5824.}
\tablehead{ 
\colhead{ID} &  $T_{\rm eff}$ &  log~g      &	$v_{\rm t}$	& [Fe/H]  &  [Mg/Fe]  & [Al/Fe]\\
             &       (K)        &   &  (\kms)  & (dex)  &  (dex) &  (dex)   }
\startdata 
\hline   
  8953    & 4578  & 1.4 & 1.8	&  --2.13$\pm$0.08  &  0.37$\pm$0.07  &  $<$0.02    \\  
 10222    & 4955  & 1.7 & 1.7	&  --2.14$\pm$0.07  &  0.51$\pm$0.10  &  $<$0.25    \\  
 10647    & 4265  & 0.7 & 2.0	&  --2.06$\pm$0.07  &  0.52$\pm$0.06  & $<$--0.03    \\  
 10967    & 4220  & 0.7 & 2.0	&  --2.15$\pm$0.05  &  0.47$\pm$0.06  &  0.62$\pm$0.04  \\  
 11276    & 4787  & 1.5 & 1.8	&  --2.20$\pm$0.07  &  0.68$\pm$0.08  &  $<$0.32    \\  
 11730    & 4382  & 1.0 & 1.9	&  --2.20$\pm$0.06  &  0.41$\pm$0.06  & $<$--0.09    \\  
 12035    & 4981  & 1.8 & 1.7	&  --2.25$\pm$0.07  &  0.46$\pm$0.10  &  $<$0.50    \\	
 12077    & 4415  & 1.0 & 1.9	&  --2.29$\pm$0.07  &  0.07$\pm$0.06  &  1.25$\pm$0.06  \\  
 12898    & 4747  & 1.8 & 1.7	&  --1.98$\pm$0.07  &  0.33$\pm$0.08  &  $<$0.42    \\  
 13068    & 4846  & 1.8 & 1.7	&  --2.10$\pm$0.07  &  0.26$\pm$0.09  &  $<$0.54    \\  
 13705    & 4707  & 1.5 & 2.1	&  --2.30$\pm$0.06  &  0.49$\pm$0.07  &  $<$0.32    \\  
 13793    & 4532  & 1.3 & 1.8	&  --2.14$\pm$0.07  &  0.48$\pm$0.07  &  $<$0.01    \\  
 13894    & 4420  & 1.1 & 1.9	&  --2.16$\pm$0.07  &  0.51$\pm$0.06  & $<$--0.01    \\  
 14000    & 4408  & 1.1 & 1.9	&  --2.18$\pm$0.07  &  0.31$\pm$0.06  &  1.26$\pm$0.06  \\  
\hline
\enddata 
\tablecomments{$~~~~~$This table is 
available in its entirety in machine-readable form.}
\label{info3}
\end{deluxetable}


{}

\end{document}